\documentclass[fleqn,10pt]{wlscirep}
\usepackage[utf8]{inputenc}
\usepackage[T1]{fontenc}
\usepackage[caption=false]{subfig}
\usepackage{listings}
\title{Physics-informed machine learning with differentiable programming for heterogeneous underground reservoir pressure management}

\author[1,2,*]{Aleksandra Pachalieva}
\author[2]{Daniel O'Malley}
\author[2]{Dylan Robert Harp}
\author[2]{Hari Viswanathan}
\affil[1]{Center for Non-Linear Studies, Los Alamos National Laboratory, Los Alamos, 87545 NM, USA}
\affil[2]{Earth and Environmental Sciences Division, Los Alamos National Laboratory, Los Alamos, 87545 NM, USA}

\affil[*]{apachalieva@lanl.gov}


\begin{abstract}
Avoiding over-pressurization in subsurface reservoirs is critical for applications like CO$_2$ sequestration and wastewater injection. Managing the pressures by controlling injection/extraction are challenging because of complex heterogeneity in the subsurface. The heterogeneity typically requires high-fidelity physics-based models to make predictions on CO$_2$ fate. Furthermore, characterizing the heterogeneity accurately is fraught with parametric uncertainty. Accounting for both, heterogeneity and uncertainty, makes this a computationally-intensive problem challenging for current reservoir simulators. To tackle this, we use differentiable programming with a full-physics model and machine learning to determine the fluid extraction rates that prevent over-pressurization at critical reservoir locations. We use DPFEHM framework, which has trustworthy physics based on the standard two-point flux finite volume discretization and is also automatically differentiable like machine learning models. Our physics-informed machine learning framework uses convolutional neural networks to learn an appropriate extraction rate based on the permeability field. We also perform a hyperparameter search to improve the model's accuracy. Training and testing scenarios are executed to evaluate the feasibility of using physics-informed machine learning to manage reservoir pressures. We constructed and tested a sufficiently accurate simulator that is 400\,000 times faster than the underlying physics-based simulator, allowing for near real-time analysis and robust uncertainty quantification.
\end{abstract}
\begin{document}

\flushbottom
\maketitle
\thispagestyle{empty}

\section*{Introduction}
\label{sec:intro}
Reservoir pressure management is essential for injection/extraction operations in the subsurface for resource extraction, carbon sequestration, climate change mitigation, and renewable energy. However, when mishandled, the pressure management strategy can cause induced seismicity, examples of which are the seismic events due to large-scale wastewater re-injection in central Oklahoma \cite{zoback2012managing,keranen2014sharp,mcnamara2015earthquake}. Another example of a failed pressure management strategy is the seismic events that followed the injection of large quantities of water at high pressure at the geothermal reservoir in Basel, Switzerland \cite{baer2007earthquakes,deichmann2008earthquakes,dyer2008microseismic}. The incident led to public distrust and, finally, the cancellation of the enhanced geothermal systems (EGS) project altogether \cite{deichmann2009earthquakes}, which was just one out of multiple EGS projects canceled due to induced seismicity concerns \cite{majer2007induced}. 

Another application that can benefit from a better understanding of pressure management is Geologic CO$_2$ sequestration (GCS). GCS at a large scale is necessary to reduce anthropogenic CO$_2$ emissions enough to combat global warming and climate change. 
%\textcolor{red}{Are we going for better understanding or speeding up and better uncertainty quantification?}
To do this efficiently, it is crucial to choose a well-suited reservoir in which 99\% of the geologically sequestered carbon will remain sequestered for over 1000 years \cite{metz2005ipcc}. This requires a pressure management strategy that minimizes the risk of leakage and potential induced seismicity through wells, faults, and fractures \cite{buscheck2011combining,cihan2015optimal,harp2017development}. This requires solving complex physics models with sufficient fidelity and enough realizations to understand and forecast the system behavior. This is only feasible if we use methods to speed up the process while keeping the results accurate.

An overwhelming amount of research is focused on using machine learning (ML) to improve the efficiency and accuracy of subsurface energy-related fluid-flow applications \cite{viswanathan2022fluid}. Machine learning has been used to create reduced order models for geologic CO$_2$ sequestration \cite{chen2018geologic,menad2019predicting,wang2020inferring,amar2020prediction}, CO$_2$ enhanced oil recovery, geothermal energy \cite{krasnov2017machine,chen2019characterization,you2020development,you2020machine}, geothermal energy \cite{li2017machine,rezvanbehbahani2017predicting,holtzman2018machine,tut2020prediction}, and oil and gas extraction \cite{hegde2017use,hanga2019machine}. For each of these applications, an effective pressure management strategy is required to mitigate the risks associated with injection/extraction operations \cite{axelsson2005sustainable,buscheck2011combining,zoback2012managing,weingarten2015high, cihan2015optimal,harp2017development}. However, the use of machine learning to manage the pressures in a subsurface injection scenario has not been investigated in depth.
 
Modeling reservoir pressure management is challenging considering the complex heterogeneity of the reservoirs and the uncertainties of the systems' input parameters. This complex heterogeneity typically requires high-fidelity physics-based models to make CO$_2$ predictions. Furthermore, characterizing the heterogeneity accurately is fraught with parametric uncertainty. Accounting for both heterogeneity (which contributes to the complexity and high computational cost of the physics model) and uncertainty demands many realizations. Performing many realizations makes this a computationally-intensive problem that is challenging for current reservoir simulation workflows. For some applications, such as the oil and gas industry, tens of thousands of wells have been drilled, resulting in large amounts of data and the development and usage of data-driven machine learning models \cite{cao2016data,mohaghegh2017data,hajizadeh2019machine}. However, this is not the case for applications such as GCS, where few wells are in use, data is scarce, expensive, and time-consuming to obtain \cite{claprood2012workflow,mishra2013maximizing}. For such data-limited applications, physics constraints are often introduced into the machine learning algorithms to regularize the neural networks training and thus augment the lack of data \cite{yang2019adversarial,meng2020ppinn}. This approach is called physics-informed neural networks (PINN) \cite{raissi2019physics}. A limitation of the PINN approach is that if the physics are not trustworthy -- there are no guarantees that the computation will quickly (or at all) converge to the correct solution. An incorrect solution could misguide the pressure management machine learning model. A major limitation of most traditional numerical models is the calculation of parameter gradients from high-fidelity physics reservoir simulations. Most fluid and transport simulators, which can simulate subsurface fluid injection/extraction rely on finite-difference gradients to evaluate many physics-based parameters \cite{pruess1991tough2,white1997stomp,harbaugh2005modflow,zyvoloski2007fehm,cmg2018advanced}. This is why such simulators are built without the use of differentiable programming (DP) and automatic differentiation (AD) techniques which are standard in machine learning approaches such as PINNs. Computing finite-difference gradients for highly-dimensionalized models (e.g., those with heterogeneous permeability fields) is computationally inefficient and often prevents the traditional physical models from being included in machine learning workflows. A solution to this problem is using DP and AD that takes advantage of the chain rule to evaluate complex derivatives more efficiently \cite{baydin2018automatic}, including for implementing trustworthy numerical models based on traditional methods such as finite difference/element/volume. 

We developed a physics-informed machine learning (PIML) framework that determines the fluid extraction rates for dedicated wells to maintain the pressure at a critical location during water injection. We consider a complex physics model that accounts for the permeability field's heterogeneity, which was overlooked or deemed too expensive by previous studies \cite{harp2021feasibility}. Instead of using an analytical single-phase model as previously done by Harp et al.\cite{harp2021feasibility}, we solved a full-physics model using the DPFEHM framework \cite{DPFEHMgit} that accounts for heterogeneity. Our physics model is differentiable, which in alternative approaches, such as PINNs, do not rigorously guarantee that the physical constraints will be satisfied. 
%\textcolor{red}{Do we want to explain the advantages of differentiable programming and what it enables us to do beyond other PINN approaches?} 
We use a steady-state equation that looks at the long-term impact of the injection/extraction. Harp et al.\cite{harp2021feasibility} considered only a fixed time in the future, which limited the ability to evaluate the sustainability of the injection process. DPFEHM allows us to combine a rigorous, trustworthy physics model with Convolutional Neural Networks (CNN) that have built-in AD. In addition, DPFEHM minimizes the time for code development by automatically creating the execution code from a short description of the equations of interest.  

One way to develop a full-physics model for pressure management in the context of CO$_2$ sequestration would require the completion of the following steps: (1) development of a simple pressure management PIML framework solving the analytical Theis model \cite{theis1935relation} as shown by Harp et al. \cite{harp2021feasibility}; (2) development of a pressure management PIML framework using a full-physics model with heterogeneous permeability field, which we demonstrate in the current study; and (3) adding a multi-phase model. Our single-phase pressure management model is more relevant to wastewater injection. However, including a full-physics model in the ML workflow and accounting for heterogeneity are essential steps towards pressure management for multi-phase injection scenarios such as CO$_2$ sequestration. Furthermore, without this PIML framework, we would be unable to do uncertainty quantification (UQ) in real-time since thousands (or more) partial differential equation-constrained optimization problems would be required.

\section{Methods}
\label{sec:methods}
Operators at underground reservoir sites require pressure management systems to make informed decisions for the injection/extraction rates to minimize the risk of leakage and induced seismicity and maximize the reservoir's performance (e.g., maximize the net fluid injected). The traditional approach to constructing pressure management systems uses full-order physics models that do not allow for UQ in real-time. Existing alternatives, such as the work of Harp et al. \cite{harp2021feasibility}, used a highly simplified homogeneous model which neglects heterogeneity. In reality, the subsurface is highly heterogeneous, so for the model to be trustworthy, heterogeneity must be accounted for. Our PIML framework uses a full-physics model and ML combined with DP, making this approach computationally and practically feasible. 

\begin{figure}
    \centering
    \includegraphics[width=\textwidth]{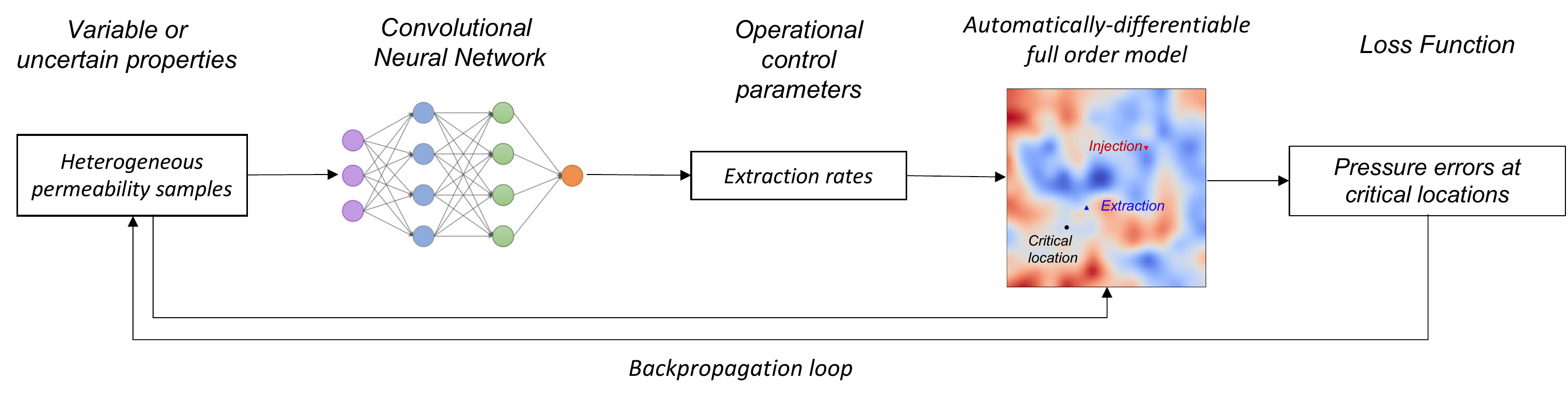}
    \caption{(Color online) Workflow diagram of physics-informed machine learning framework for managing reservoir pressures at a critical location during subsurface fluid injection. The key innovation of the surrogate model is the automatically-differentiable full-order model that allows for heterogeneity.}
    \label{fig:piml_workflow}
\end{figure}

A schematic representation of the proposed PIML framework is shown in Figure\,\ref{fig:piml_workflow}. We use this framework to determine the fluid extraction rates at an extraction well to maintain the pressure at a critical location in a reservoir with a heterogeneous permeability field. We achieve this by using a neural network model (NNM) trained on a set of permeability fields that, along with the NNM-predicted extraction rate, act as an input to the full-physics DPFEHM model. The NNM is trained to determine the extraction rates for a heterogeneous permeability field. The physics constraints needed for the training process are incorporated through the DPFEHM framework, which provides the physics information in our PIML framework. Our framework has similarities to the works of Harp et al. and Srinivasan et al. \cite{srinivasan2021machine,harp2021feasibility}, that use physics-informed neural networks with physical constraints used during the NNM training. A significant difference in our approach is that instead of using a differentiable, simple analytical solution, we use a differentiable numerical physics model. Alternative approaches such as PINNs do not have rigorous guarantees that the physical constraints will be satisfied.

We executed a suite of training scenarios using a single-phase model with heterogeneous permeability fields randomly generated using a Gaussian distribution function. Since the model has many parameters, computing finite-difference gradients become infeasible, and the only efficient approach to solve the problem is by using reverse-mode automatic differentiation. We perform a hyperparameter search by varying the batch size (i.e., the number of training samples included in each gradient calculation) and the learning rate (i.e., the step size controlling the rate at which each iteration moves towards the minimum of the loss function). 

\subsection*{Physics model}
\label{subsec:phys_model}
We consider the pressure change during injection/extraction of a 
single-phase fluid flow in a heterogeneous permeability field. Such subsurface reservoirs are modeled using the following equation
\begin{equation}
    \nabla\cdot\big(k(x)\cdot\nabla h\big) = f, 
    \label{eq:physics_model}
\end{equation}
where $k(x)$ are heterogeneous permeability fields depending on the position $x$, $h$ is the pressure head and $f$ is the injection/extraction. This is a steady-state equation which allows us to evaluate the long-term impact of the injection and the extraction on the pressure head. 

The equation is solved using DPFEHM \cite{DPFEHMgit}. DPFEHM uses the standard two-point flux finite volume approximation, which results in a trustworthy solution to the physics model. In addition, DPFEHM has built-in AD that makes the transition between the physics model and the machine learning model seamless. 

\subsection*{PIML framework}
\label{subsec:piml_model}
As shown in Figure\,\ref{fig:piml_workflow}, our PIML framework trains an NNM to estimate the extraction rates at dedicated extraction wells to maintain predefined pressures at critical locations during fluid injection. This is necessary near faults with high induced seismicity risk or at abandoned wellbores with leakage potential. To make the model more realistic, we randomly initialized heterogeneous permeability fields using a Gaussian distribution function. The PIML framework is part of the DPFEHM package, which is available at \url{https://github.com/OrchardLANL/DPFEHM.jl}.
\vspace{0.25cm}

The PIML workflow consists of the following steps:
\vspace{0.1cm}
\begin{enumerate}
    \item Generate a training dataset that consists of $N_b$ batches and $N_s$ samples per batch. We use heterogeneous permeability samples, randomly initialized using a Gaussian distribution function.
    \item Construct a CNN with an input layer that accepts a permeability field and an output layer that estimates the fluid extraction rates at the extraction well.
    \item Calculate a loss function that quantifies the error between the model's overpressure and the target overpressure at a critical location. 
    \item Train the CNN to determine the extraction rates that minimize the error between the model's overpressure and the target overpressure at a critical location by adjusting the CNN model parameters based on the loss-function parameter gradients. 
\end{enumerate} 

In step 2, the PIML framework trains a CNN based on LeNet-5 \cite{lecun1989back} to determine the extraction rates at an extraction well to maintain pressure at critical locations, such as faults with induced seismicity risk or abandoned wellbores with leakage potential, during fluid injection at dedicated injection wells with heterogeneous permeability fields. LeNet-5's architecture consists of a convolutional encoder with two convolutional layers and a dense block with three fully-connected layers. For this work, we use a slightly modified LeNet-5 CNN with the following architecture:
\begin{lstlisting}[language=Python]
model = Chain(Conv((5, 5), 1=>6, relu),
              MaxPool((2, 2)),
              Conv((5, 5), 6=>16, relu),
              MaxPool((2, 2)),
              flatten,
              Dense(1296, 120, relu),
              Dense(120, 84, relu),
              Dense(84, 1)) |> f64
\end{lstlisting}

We use two convolutional layers, two subsampling layers, a flattening operation, and a fully connected dense block. Each convolutional layer uses a 5x5 kernel and a max pooling subsampling layer which calculates the maximum of the values present in each kernel. The first convolutional layer has 6 output channels, and the second one has 16. The activation function with a stride of 2 reduces the dimensionality through downsampling by a factor of 4. We flatten the output of the convolutional block by taking the four-dimensional input and transforming it into two-dimensional input to be compatible with the fully-connected dense block. The dense block has three fully-connected layers, with 120, 84, and 1 outputs, respectively. Relu activation functions are used for all hidden layers as $\sigma(x) = \max(0,x)$, with $x$ being the input and $\sigma(x)$ being the output of the neuron. The last dense layer outputs a single value associated with the extraction rate, $Q_\mathrm{ext}$. 

During preliminary investigations, we explored the use of more complex deep neural networks (such as the VGG16, \cite{simonyan2014very}) with a larger number of neurons and/or more hidden layers; however, they did not improve the convergence speed and thus the performance of the PIML framework. On the contrary, in some cases, they led to slower convergence and worse overall results. The reason for this is most likely that larger networks are more prone to complex response surfaces, which increases the chance for the training to become trapped in local minima. By contrast, our empirical results suggest that LeNet-5 provides a good balance between strong performance and ease of training. As we transition to more complex physics models, the ease of training becomes more critical when the data sets become smaller. Nonetheless, the optimal architecture for these problems remains an exciting area for research.

In step 3, the loss function is defined as the sum of the squared errors between the simulated and the target overpressures at a critical location as 
\begin{equation}
    \mathcal{L}(\theta) = \sum_i^{N_b} \sum_j^{N_s}\Big[\Delta h\left(Q_\mathrm{NN}\big(\theta, k_j(x)\big), k_j(x)\right)-\Delta h^\mathrm{target}\Big]^2,
    \label{eq:loss}
\end{equation}
with $N_b$ being the number of batches, $N_s$ the number of samples per batch, and $\Delta h^\mathrm{target}$ the target overpressure. The predicted overpressure $\Delta h$ is a function of the injection rate $Q_\mathrm{NN}$ and the permeability $k_j(x)$ at a specific position. $Q_\mathrm{NN}$ is calculated using the LeNet-5 CNN model and takes two parameter, $\theta$ being the model parameters and $k_j(x)$ permeability. From the loss function we calculate the root-mean-square error (RMSE) as
\begin{equation}
    \mathrm{RMSE} (\theta) = \sqrt{\frac{\mathcal{L}(\theta)}{N_b N_s}}.
    \label{eq:rmse}
\end{equation}

In step 4, the NNM is trained to minimize the loss function $\mathcal{L}(\theta)$ as 
\begin{equation}
    \min_{\theta}=\mathcal{L}(\theta).
    \label{eq:min}
\end{equation}
using an Adam optimizer from the Julia Flux package \cite{innes2018flux}.
We use the DPFEHM package to solve the full-physics model as shown in Eq.\ref{eq:physics_model}. Within the DPFEHM framework, we automatically differentiate the physics model using Julia's Zygote package \cite{innes2018don}. 

\subsection*{Differentiable programming}
\label{subsec:diff_prog}

Traditional physics models are rarely automatically differentiable and require the usage of finite difference methods to compute parameter gradients \cite{pruess1991tough2,white1997stomp,harbaugh2005modflow,zyvoloski2007fehm,cmg2018advanced}. This is computationally inefficient and makes incorporating these models in a machine learning workflow infeasible when the number of parameters is large. When looking at a large number of physics-model parameters in a PIML framework, choosing the most efficient way to calculate gradients is essential for the algorithm's success. This is because computing the gradient of the loss function is central to training machine learning models.

In computational fluid dynamics, finite-differences or numerical differentiation are often used to compute gradients, but that could be computationally expensive, especially when the number of input parameters grows. The number of model runs required to compute the gradient is proportional to the number of input parameters. When using finite difference (FD), the quality of the solution is also greatly influenced by the truncation and round-off errors associated with different finite difference formulas. An alternative is using differentiable programming (DP), and in particular, reverse-mode AD, which utilizes the chain rule to calculate complex derivatives. It breaks a computer program down into elementary operations, which allows the derivatives to be evaluated accurately to working precision. Reverse-mode AD becomes more efficient when the number of model inputs is large (e.g., the parameters of a neural network) and the number of model outputs is small (e.g., the loss), which can be demonstrated with the following example: The LeNet-5 model uses approximately 1 million model parameters, which gradients can be obtained using AD in a single forward and backward model pass regardless of the number of model parameters; in comparison, using a central finite-difference model, we would have to calculate two forward simulations per model parameter which results in 2 million forward calls. 
In AD, the advantage of using reverse-mode AD (akin to adjoint methods) comes from its computational cost being independent of the number of design variables. By contrast, the cost of the forward sensitivity analysis increases linearly with the number of design variables \cite{giles2005using}. Depending on the simulator and the application, the AD adjoint simulations can be more expensive than the forward simulations; however, the discrepancies in execution time have been decreasing with the development of DP \cite{baydin2018automatic,innes2019differentiable}. For a physics simulator, which typically involves a solution of a system of equations, the backward pass requires only the solution of a linear system of equations (even if the underlying equations are nonlinear). So in this setting, the backward pass is rarely more expensive than the forward pass.

\section*{Results}
\label{sec:results}
\begin{figure}
    \centering
    \includegraphics[width=0.55\textwidth]{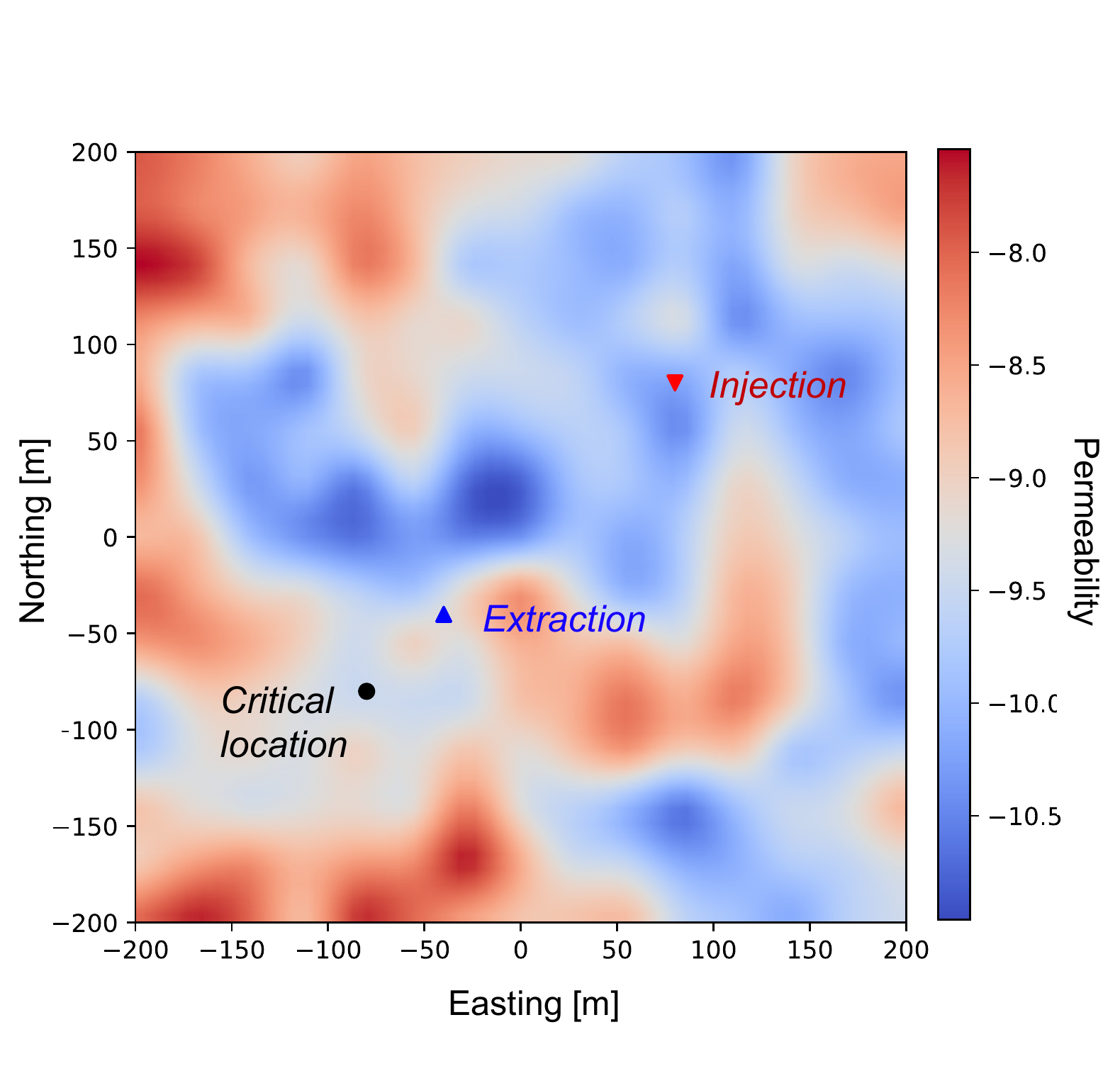}
    \caption{(Color online) Diagram of a single-phase pressure management with heterogeneous permeability field including one injection and one extraction wells, and a critical location.}
    \label{fig:sim_config}
\end{figure}

%\subsection{Scenario configuration}
The physics model includes an injection well, an extraction well, and a critical location. The injection well injects water at a rate of $1.0\,\mathrm{MMT/y}$ (million metric tons per year), which is equivalent to $0.031688\,\mathrm{m}^3/\mathrm{s}$. The extraction well protects the critical location where a target overpressure (change in pressure from pre-injection conditions) is set to 0.0 MPa. %$0.0$ mmH2O
The permeability field is randomly initialized using a multivariate Gaussian distribution function with a correlation length of $50\,$m and log standard deviation equal to 1. A schematic representation of the simulation configuration is shown in Figure\,\ref{fig:sim_config}, where we indicate the positions of the injection, extraction, and critical location. The permeability field is illustrated using a color map, where blue and red denote low and high permeability, respectively. The size of the simulated domain is 400 meters on each side. Our PIML framework trains the neural network to achieve the target overpressure at the critical location for a given permeability field. We solve a steady-state equation that captures the long-term impact of the injection/extraction. The background reservoir pressure in our simulation is equal to 19.0\,MPa, which is in line with MPC 26-5 well located in Kemper County, Mississippi. MPC 26-5 well is part of the ECO$_2$S project and its depth is 1791 meters \cite{duguid2018co2,riestenberg2018survey}. 

The PIML algorithm uses 512, 1024, 2048, or 4096 training and disjoint testing samples per epoch depending on the batch size, where the batch size ranges from 32 to 256. The testing data is sampled only once from a multivariate Gaussian distribution, while the training data is sampled in every epoch. That is, the model never sees the same data point twice, so there is little risk of overfitting, a unique characteristic of the DP that allows for data to be generated on the fly compared to other ML models that use just a fixed set of data. This is possible because our DP approach has the physics model in the training loop. Therefore, generating the training data and running the model is unnecessary before the training begins. The total number of executed epochs is 10,000, but similar results could be obtained using a much smaller number of epochs on the order of 4,000. The training samples are randomly initialized before training each epoch, making the total number of unique training samples for the batch size of 256 equal to 40,960,000 throughout the ML training. The test samples are initialized only once before starting the first epoch. We ran the training on a central processing unit (CPU) with an AMD EPYC 7702P 64-Core processor without parallelization paradigms. 

We perform a hyperparameter search in which we test the RMSE for a variety of batch sizes ($\mathrm{bsize}\in[32, 64, 128, 256]$), and learning rates ($\mathrm{lr}\in[1\mathrm{e}^{-3}, 3\mathrm{e}^{-4}, 1\mathrm{e}^{-4}]$) as shown in Table\,\ref{tbl:hyperparamerter_search}. The hyperparameter search shows that by decreasing the learning rate from $\mathrm{lr}=1\mathrm{e}^{-3}$ to $\mathrm{lr}=1\mathrm{e}^{-4}$, the RMSE decreases three to four times for each batch size. The best results are achieved using a batch size and learning rate equal to $256$ and $1\mathrm{e}^{-4}$, respectively. The results can be potentially improved by further decreasing the learning rate. However, we did not expand the search since the model was producing results that were within acceptable limits.

\begin{table}[ht]
\centering
\begin{tabular}{|l|cccc|}
\hline
 & $\mathrm{bsize} = 32$ & $\mathrm{bsize} = 64$ & $\hspace{0.2cm} \mathrm{bsize} = 128$ & $\mathrm{bsize} = 256$ \\ \hline
 $\mathrm{lr} = 1\mathrm{e}^{-3}$ & 0.018658 & 0.013856 & 0.010190 & 0.009823 \\
 $\mathrm{lr} = 3\mathrm{e}^{-4}$ & 0.005779 & 0.015159 & 0.005228 & 0.006147 \\
 $\mathrm{lr} = 1\mathrm{e}^{-4}$ & 0.005332 & 0.003719 & 0.002852 & 0.002786 \\
 \hline
\end{tabular}
\caption{\label{tbl:hyperparamerter_search}Hyperparameter search to determine the optimal ML parameters. Minimum testing RMSE for a variety of batch sizes, $\mathrm{bsize}\in[32, 64, 128, 256]$; and learning rates, $\mathrm{lr}\in[1\mathrm{e}^{-3}, 3\mathrm{e}^{-4}, 1\mathrm{e}^{-4}]$. The results are given in pressure units [MPa].}
\end{table}

\begin{figure}
    \centering
    \includegraphics[width=1.\textwidth]{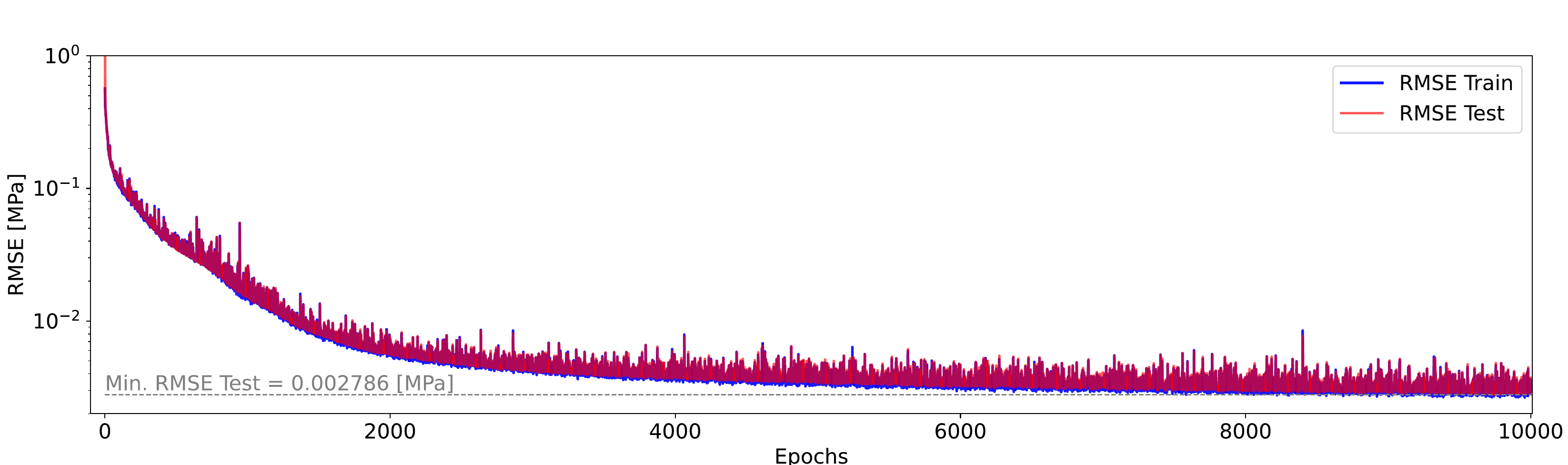}
    \caption{(Color online) RMSE for the best training set from the hyperparameter search ($\mathrm{bsize} = 256$, and $\mathrm{lr} = 1\mathrm{e}^{-4}$). The minimum RMSE is equal to 0.002786\,MPa overpressure at the critical location.}
    \label{fig:loss_results}
\end{figure}
Figure\,\ref{fig:loss_results} illustrates the rate at which the RMSE decreases with the number of epochs. The results are for the best hyperparameters we tested during PIML training, which uses a batch size of 256 and a learning rate equal to $1\mathrm{e}^{-4}$. The figure shows the training RMSE error in blue and the testing RMSE in red. We observe a rapid decrease in the RMSE at the beginning of the training. In later epochs (after approximately 4000 epochs), the RMSE starts to decrease much slower, reaching a plateau around 8000 epochs with an overall error reduction of 99\%. The minimum RMSE reached using the testing data samples is $0.002786$\,MPa, which is small in comparison to the overpressure caused by the injection and makes the model sufficiently accurate to be used in practice. That is, other factors beyond errors in the injection rate, such as uncertainty about the heterogeneity, will impact the decision-making process for the extraction rate, and this model could be used to help find an appropriate extraction rate given uncertainty by predicting an ensemble of extraction rates for a variety of permeability fields.

\begin{figure}
     \centering
     \subfloat[]{\label{subfig-1:results_totalpressure}{\includegraphics[width=0.53\textwidth]{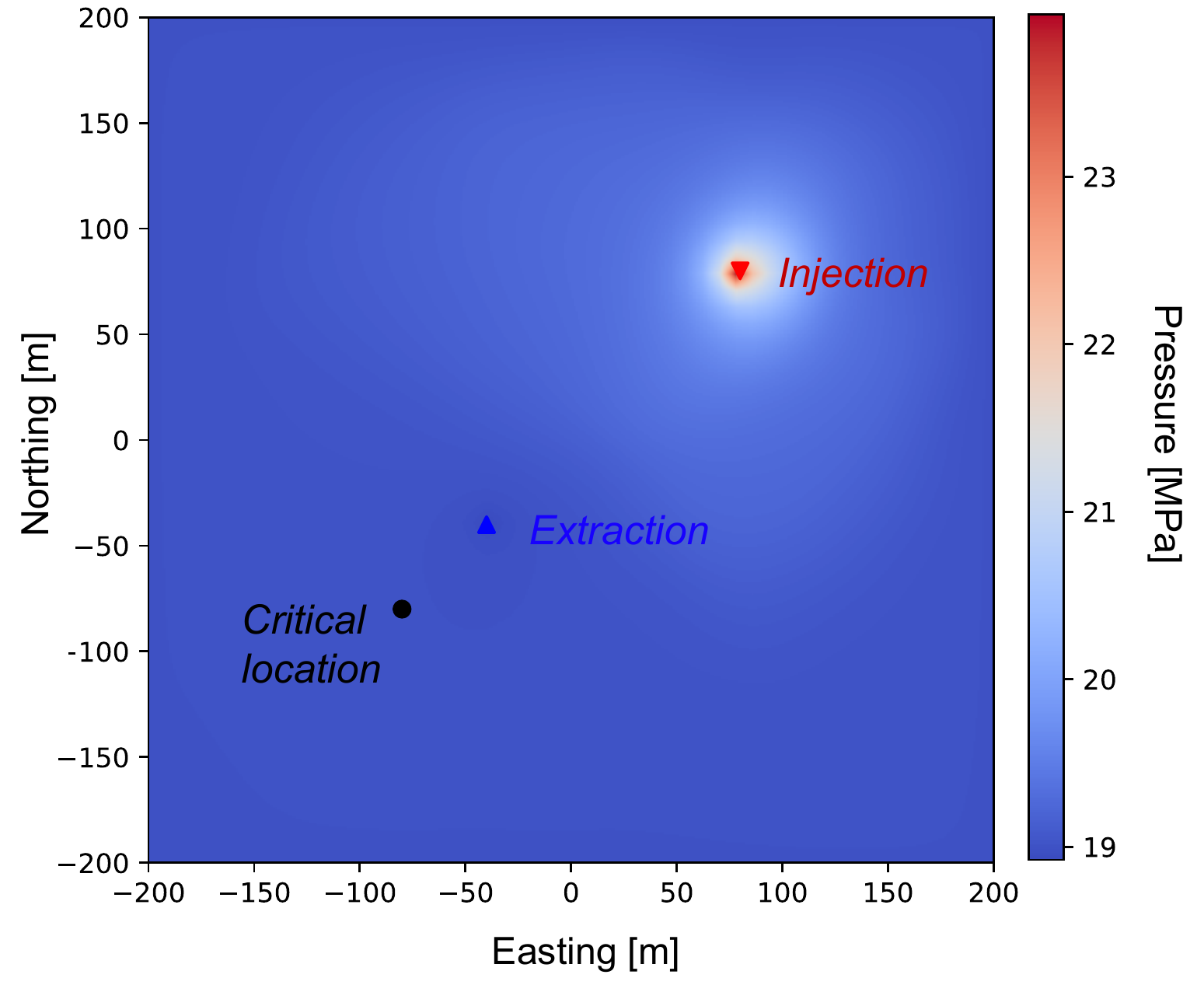}}}
     \subfloat[]{\label{subfig-2:results_overpressure}{\includegraphics[width=0.48\textwidth]{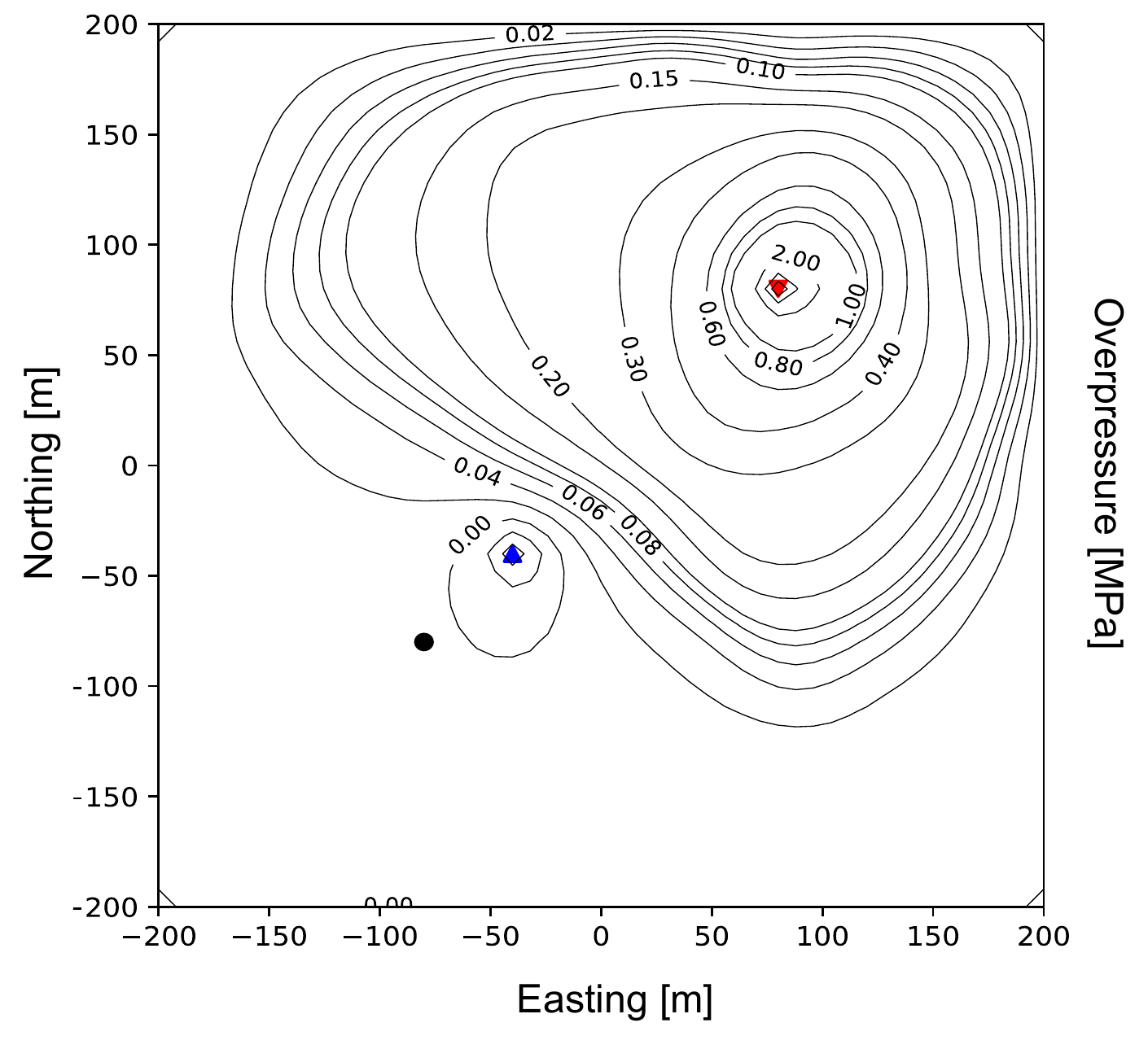}}}
    \caption{(Color online) (a) Resulting pressure distribution within the reservoir. The positions of the injection/extraction and critical locations are indicated; (b) Overpressure values within the prescribed domain extracted from the PIML framework. The blue and red triangles denote respectively the position of the extraction and the injection wells, respectively, while the circle depicts the critical location at which a target pressure is set.}
    \label{fig:results_pressure}
\end{figure}

After training the PIML frameworks, we can use them to generate extraction rates and overpressures at the critical location given a heterogeneous permeability field. In Figure\,\ref{fig:results_pressure}, we illustrate the resulting pressures and overpressures for a random permeability field within the domain. We used the same permeability field as shown in Figure\,\ref{fig:sim_config}, and the PIML framework for which we achieved the lowest RMSE ($\mathrm{bsize}=256$ and $\mathrm{lr}=1\mathrm{e}^{-4}$). The background pressure of the MPC 26-5 well is 19.0\,MPa, which is reflected in the total pressure of the well depicted in Figure\,\ref{subfig-1:results_totalpressure}. Figure\,\ref{subfig-2:results_overpressure} shows that the prescribed pressure at the critical location, 0.0\,MPa, is obtained by the PIML framework with good accuracy and, despite the small deviation, still falls into an acceptable range. As expected, the overpressure rates at the injection well are considerably larger, up to 6\,MPa. At the same time, at the extraction well, we achieve negative overpressure, which protects the critical location from exceeding the prescribed 0.0\,MPa. 

\begin{figure}
     \centering
     \subfloat[]{\label{subfig-1:extraction_rate}{\includegraphics[width=0.5\textwidth]{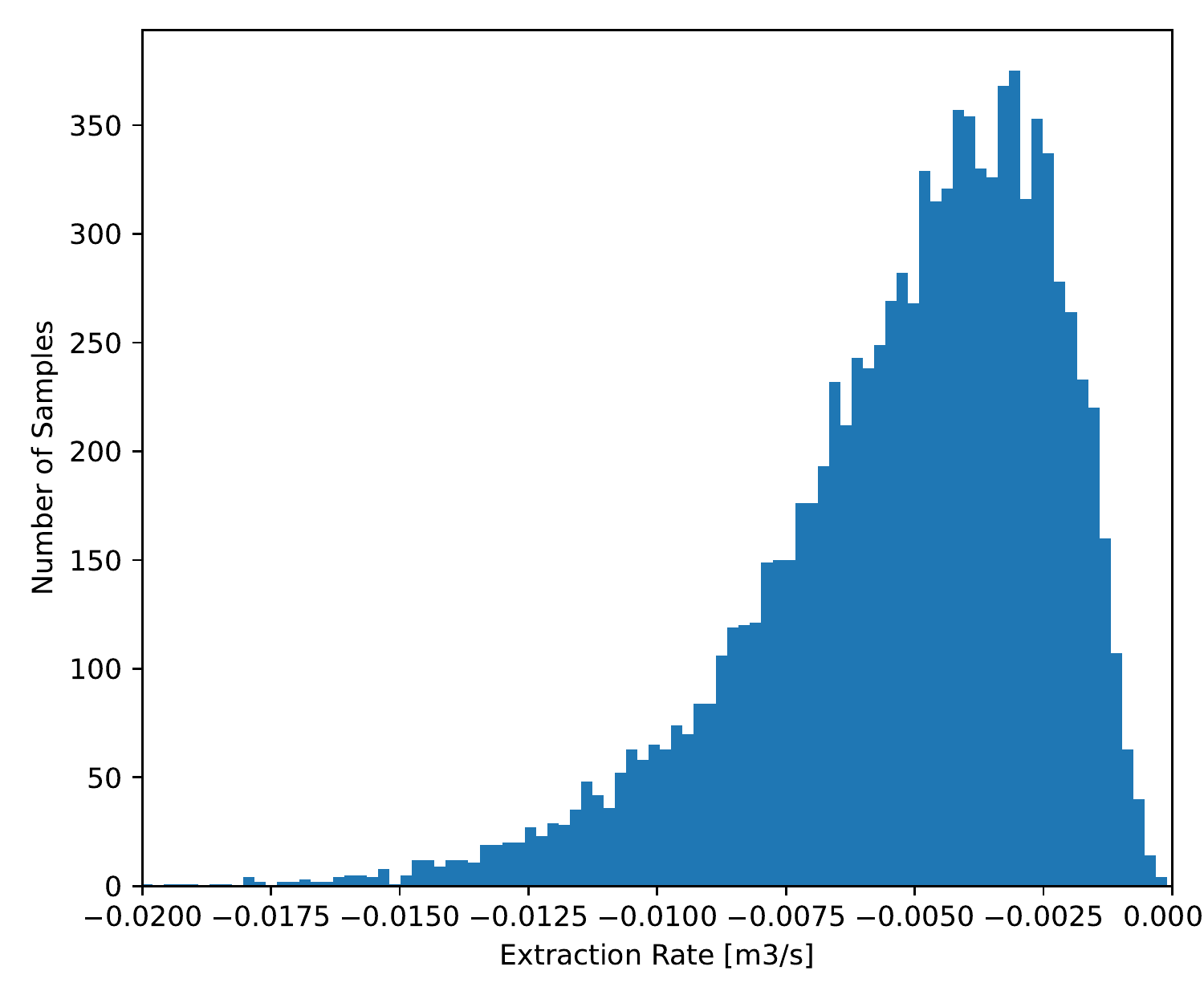}}}
     \subfloat[]{\label{subfig-2:overpressure}{\includegraphics[width=0.5\textwidth]{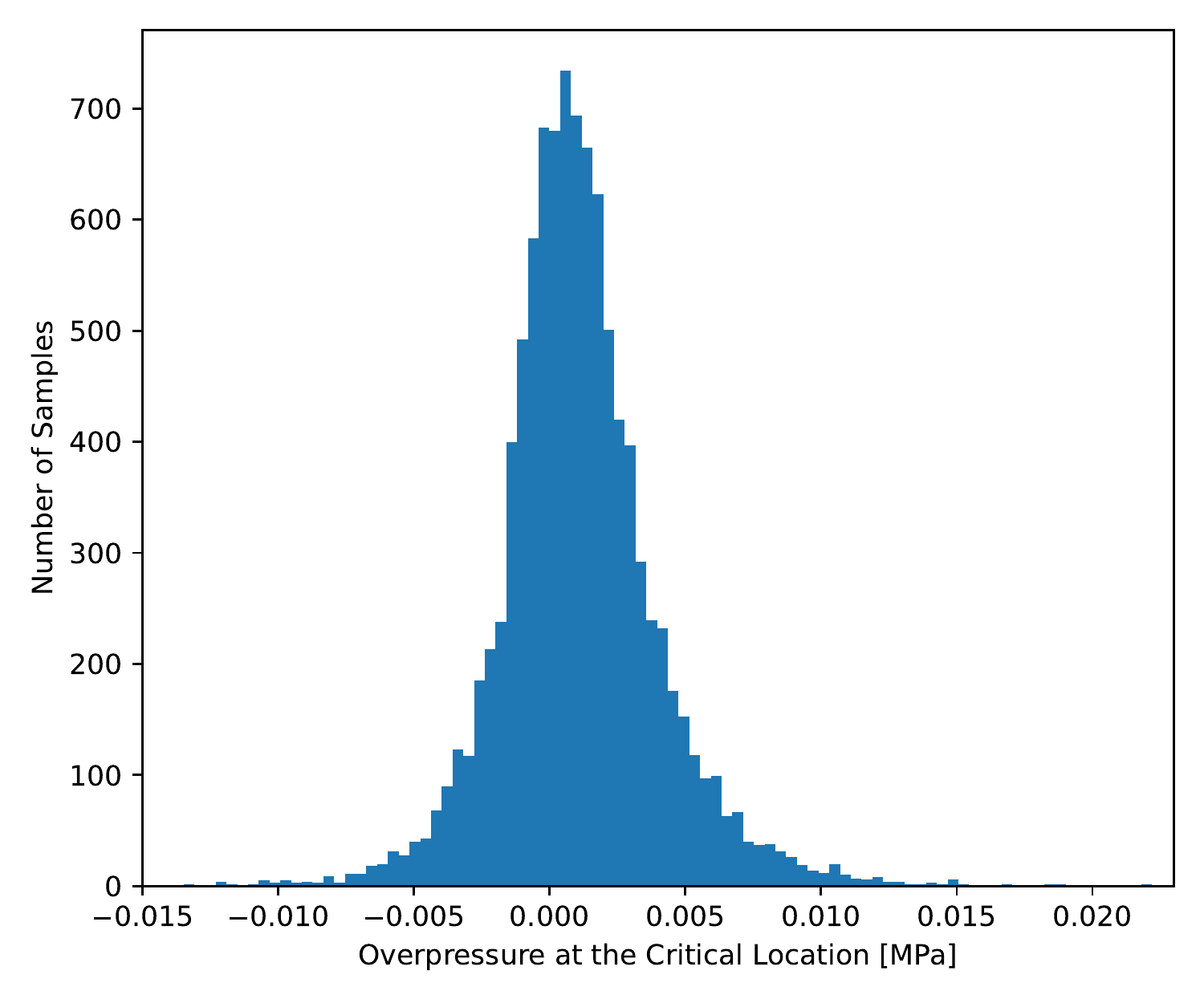}}}
     \caption{(Color online) (a) Extraction rates in cubic meters per second [$\mathrm{m}^3/\mathrm{s}$]; (b) Overpressure at the critical location in pressure units [MPa].}
    \label{fig:distributions}
\end{figure}

Next, we look at the distribution of extraction rates and overpressures at the critical location after running 10,000 random heterogeneous field samples through our PIML framework as shown in Figure\,\ref{fig:distributions}. The distribution function of the extraction rate, depicted in Figure\,\ref{subfig-1:extraction_rate}, is skewed and shows that most of the samples require an extraction rate in the range of $-0.009$\,$\mathrm{m}^3/\mathrm{s}$ to $-0.001$\,$\mathrm{m}^3/\mathrm{s}$ in 90\% of the time. Compared to our injection rate of 1.0\,MMT/year, the extraction rate is equal to $-0.283824$\,MMT/year, and $-0.031536$\,MMT/year, respectively. Figure\,\ref{subfig-2:overpressure} illustrates that the overpressure at the critical location deviates very slightly from the prescribed overpressure, specifically in the range of $-0.003$\,MPa to $0.006$\,MPa in 90\% of the time. These results confirm that the PIML framework can provide useful information for the operators of pressure management systems.

Overall, solving the physics model on a CPU (AMD EPYC 7702P 64-Core) takes $5.9\mathrm{e}^{-3}$\,seconds, and obtaining the extraction rates using the trained PIML model on a GPU (NVIDIA RTX A6000) takes only $1.4\mathrm{e}^{-8}$\,seconds (equivalent to 14 nanoseconds) per sample. Our framework solves the constrained optimization model more than 400\,000 times faster than solving the physics model allowing for near real-time analysis and robust uncertainty quantification. We obtained these results using a total number of 50\,000 to fully utilize the GPU.

\section*{Discussion}
\label{sec:discussion}

We have demonstrated that the PIML framework can successfully manage subsurface pressures by controlling fluid extraction in heterogeneous permeability fields. Such operations are relevant for resource extraction (oil and gas), climate mitigation (CO$_2$ sequestration), and production of renewable energy (geothermal energy). A similar PIML approach has already been evaluated by \cite{harp2021feasibility} using homogeneous permeability. We have added a significantly more complex physics model using the DPFEHM framework to ensure a seamless transition between the execution of the physics model and the machine learning model. We show that a PIML framework with built-in AD can train an NNM to determine the fluid extraction needed to achieve pressure management goals while injecting large amounts of fluid into the subsurface. This has the potential to help operators at sites effectively manage reservoir pressures and can be coupled with decision-making strategies to attain operational efficiencies \cite{martin1967bayesian,charnes1978measuring,geng2001intelligent,ben2006info,zhang2013decision}.

The computational cost is modest enough that it can be run on a single CPU with AMD EPYC7702P 64-Core processor without parallelization. However, the full-physics model and the heterogeneity add complexity to the model that is only feasible because of the DPFEHM framework and its built-in AD. The number of training epochs can be reduced depending on the accuracy demands, which can be a helpful strategy when considering a more complex physics model. Our results showed that, though we trained for 10,000 epochs, a much smaller number could be used to obtain accuracy consistent with realistic operational goals. This could be important when using more complex physics models (such as multi-phase flow) to keep the computational cost manageable. Parallelization could also be exploited in these cases.

We tested different convolutional neural networks, one of them was the VGG Net, a very deep CNN often used for image recognition \cite{simonyan2014very}. To execute the VGG Net more efficiently, we used a hybrid approach, in which the ML model was executed on the GPU while the physics model was executed on the CPU. Even though the VGG Net is significantly deeper, it did not improve the RMSE of the training and sometimes performed worse. However, such a neural network could be more efficient when training on even more complex physics models such as multi-phase models needed for investigating climate mitigation and, in particular CO$_2$ sequestration. These models might require a more complex network to deal with the more complex physics present in these cases, but this is a topic for future research.

In addition, we batch-parallelized the model using Julia's parallel map operation implemented as a \texttt{pmap}-function. Since our physics model is only modestly expensive, the communication between processes was longer than the time needed for the computation. Therefore, the \texttt{pmap} did not improve the performance. However, this technique could be used more effectively when investigating more complex physics models, such as multi-phase flow. Again, this is a topic for future research.

\section*{Conclusions}
We applied a PIML framework to a subsurface pressure management problem that considers the pressure change during injection/extraction. We considered a single-phase steady-state fluid flow with heterogeneity that looks at the long-term impact of the injection/extraction on the reservoir. In our PIML framework, a convolutional NNM is trained to determine fluid extraction rates at a dedicated critical reservoir location during the injection. We performed a  hyperparameter search, which showed that decreasing the learning rate while increasing the batch size improves the results. In conclusion, we would emphasize the following observations:
\label{sec:conclusion}
\begin{itemize}
    \item A PIML framework can train a convolutional NNM to manage reservoir pressures with heterogeneous permeability fields resulting in small deviations from the target overpressure. We accomplished this by combining a differentiable full-physics simulator with a convolutional neural network. 
    \item This problem is only feasible because of the DPFEHM framework, which has a built-in AD. To solve the physics model, we use the DPFEHM framework as part of the PIML framework. 
    \item DPFEHM allows us to bridge the gap between numerical models and machine learning techniques because both are compatible with the same AD frameworks
    \item The hyperparameter search shows that decreasing the learning rate and increasing the batch size is beneficial for bringing down the RMSE of the ML training. Our results can be potentially improved by further decreasing the learning rate. 
    \item We tested a hybrid implementation of the PIML framework using a deeper convolutional neural network (VGG Net). In that case, the physics model is solved on a CPU, while the machine learning model is solved on a GPU. The results showed no improvement in the ML training, concluding that the relatively simple LeNet network suffices for this problem.
    \item We also batch-parallelized the combined NNM and physics model using Julia's parallel map operation (\texttt{pmap}). However, for the single-phase steady-state flow problem with heterogeneity, the added communication costs outweigh the benefits of parallelization. This is because the physical model, in this case, is relatively inexpensive. We anticipate that for more complex physics models, such as introducing a multi-phase flow, will benefit from such parallelization. 
    \item A natural next step in this line of research is to study this problem in the context of multi-phase flows, which are essential for CO$_2$ sequestration, among other applications.
\end{itemize}


\bibliography{references}

\section*{Acknowledgements}
AP acknowledges funding from Center for Nonlinear Studies (CNLS) at Los Alamos National Laboratory. DO acknowledges support from Los Alamos National Laboratory's Laboratory Directed Research and Development Early Career Award (20200575ECR). The authors acknowledge support from the US Department of Energy's Science-informed Machine-learning for Accelerating Real-Time decisions (SMART) initiative.

\section*{Author contributions statement}
Aleksandra Pachalieva: Data curation, Formal analysis, Investigation, Methodology, Software, Validation, Visualization, Writing - original draft, Writing - review \& editing. Daniel O’Malley: Conceptualization, Formal analysis, Software, Methodology, Writing - review \& editing. Dylan Robert Harp: Software, Writing - review \& editing. Hari Viswanathan: Funding acquisition, Writing - review \& editing. 

\section*{Competing interests}
The author(s) declare no competing interests.

%\textcolor{red}{Must include all authors, identified by initials, for example:
%A.A. conceived the experiment(s),  A.A. and B.A. conducted the experiment(s), C.A. and D.A. analysed the results.  All authors reviewed the manuscript.}

%\section*{Additional information}

%\textcolor{red}{To include, in this order: \textbf{Accession codes} (where applicable); \textbf{Competing interests} The authors declare that they have no known competing financial interests or personal relationships that could have appeared to influence the work reported in this paper.}

%\textcolor{red}{The corresponding author is responsible for submitting a \href{http://www.nature.com/srep/policies/index.html#competing}{competing interests statement} on behalf of all authors of the paper. This statement must be included in the submitted article file.}

\end{document}